\documentclass[journal=inoraj,manuscript=article]{achemso}
\setkeys{acs}{articletitle = true,doi=true}
\usepackage[T1]{fontenc}       
\usepackage{siunitx}
\usepackage{amsmath}

\author{Haibei Huang}
\author{Willem Van den Heuvel}
\author{Alessandro Soncini}
\email{asoncini@unimelb.edu.au}
\affiliation{School of Chemistry, University of Melbourne, Melbourne, Victoria 3010, Australia}

\title[An \textsf{achemso} demo]
  {Lanthanide-radical magnetic coupling in [LnPc$_2$]$^0$:  Competing exchange mechanisms captured via \textit{ab initio} multi-reference calculations}

\abbreviations{SMMs,CASSCF,RASSCF,Ln,MOs,SOMO,AFM,FM,NMR,DFT}
\begin{document}

%
%
%
%
%
%
\begin{abstract}
We present a computational investigation of the intramolecular exchange coupling in
[LnPc$_2$]$^0$ (Ln = Tb, Dy, Ho, and Er) between the Ln$^{3+}$ 4f electrons and
the spin-1/2 radical on the phthalocyanine ligands. A series of \textit{ab
initio} multi-configurational/multi-reference Complete/Restricted Active Space
Self-Consistent-Field calculations (CASSCF/RASSCF), including non-perturbative
spin--orbit coupling, were performed on [LnPc$_2$]$^0$ and on the smaller
model compound [LnPz$_2$]$^0$. We find that the exchange coupling mechanisms
are restricted by symmetry, but also dependent on the spin polarization effect
triggered by the Pc$_2$ ligands $\pi$--$\pi^*$ excitations. The calculated
exchange splittings are small, amounting to at most a few \si{cm^{-1}}, in
disagreement with previous literature reports of strong antiferromagnetic
coupling, but in good agreement with recent EPR experiments on [TbPc$_2$]$^0$.
Furthermore, the coupling strength is found to decrease from [TbPc$_2$]$^0$ to
[ErPc$_2$]$^0$, with decreasing number of unpaired electron spins in the
lanthanide ground (Hund's rule) Russell--Saunders term.
\end{abstract}

\section{Introduction}
Single molecule magnets based on the lanthanide phthalocyanine double-decker
([LnPc$_2]^{\pm1/0}$, Ln = lanthanide, Pc = phthalocyanine) are of particular
interest due to their large barrier to magnetic relaxation
and high blocking temperatures, especially compared to traditional single
molecule magnets based on transition metals
\cite{woodruff2013lanthanide,aromi2006synthesis,wang2016single,feltham2014review,ishikawa2003lanthanide,ishikawa2008effects,demir2015radical}.

LnPc$_2$ comes in a number of oxidation states, one of the most interesting
forms being the neutral [LnPc$_2$]$^0$, partly because it has a larger
barrier for magnetic relaxation, and also because it can be
easily sublimated in ultra-high vacuum (UHV) deposition to fabricate molecular
spintronic devices
\cite{vitali2008electronic,katoh2009direct,stepanow2010spin,biagi2010x,candini2011graphene,rizzini2011coupling,vincent2012electronic,komeda2014double,Urdampilleta2015,candini2016spin,marocchi2016relay}. 
Its interest
also stems from the fact that [LnPc$_2]^{0}$ has an unpaired electron in the
Pc$_2$ ligand moieties, which has been argued to mediate the exchange coupling
between the localized Ln magnetic moment and underlying substrates, such as
magnetic thin films or carbon nanostructures
\cite{rizzini2011coupling,komeda2014double,candini2016spin,marocchi2016relay}.

Understanding the strength and nature of the exchange coupling between the Ln$^{3+}$ 4f electrons
and the organic radical delocalized over the Pc$_2$ rings thus represents an important task. The
first contribution in that direction was made in a study of the temperature-dependent magnetic
susceptibility of powder samples of [LnPc$_2$]$^0$.\cite{Trojan1991,trojan1992strong}
In that work, the authors reported saturated values of $\chi T$ which were
systematically smaller than what is theoretically expected for an independent Ln(III)--radical pair.
For example, they recorded the following values of $\chi T$ (cm$^3\,\mathrm{K}\,\mathrm{mol}^{-1}$)
at 300 K (theoretically expected in parentheses): Tb 9.2 (\textit{12.2}), Dy 13.0 (\textit{14.5}),
Ho 11.3 (\textit{14.4}), Er 8.4 (\textit{11.9}). To explain these results the authors proposed that
a strong exchange interaction must exist between the lanthanide and the Pc$_2$ radical. This
interaction should be antiferromagnetic and at least as large as room temperature ($\approx
\SI{200}{cm^{-1}}$) in order to explain the observed values, which at 300 K are appreciably lower
than expected for the uncoupled systems.\cite{Trojan1991,trojan1992strong} It should be noted that
a coupling of that magnitude is unusual for exchange involving 4f electrons. In view of the small
overlap between the highly localized 4f orbitals and the magnetic orbital(s) of the exchange
partner, a much weaker interaction is expected. This issue was not mentioned by the
authors\cite{Trojan1991,trojan1992strong}, and their conclusion that the lanthanide is strongly and
antiferromagnetically coupled to the Pc$_2$ radical has been repeated unchallenged in review
articles.\cite{Kobayashi2002,Dreiser2015,demir2015radical}

Recently, evidence to the contrary was derived from a single-crystal EPR experiment
on [TbPc$_2$]$^0$.\cite{komijani2018} The field and angle dependent resonance frequencies were found
to be consistent with a \emph{small ferromagnetic} interaction described by the Ising Hamiltonian 
\begin{equation}
  -2J_\mathrm{eff} \tilde{S}^{\mathrm{Ln}}_z S^{\mathrm{Pc}_2}_z.
\end{equation}
Here, $\tilde{S}^{\mathrm{Ln}}$ denotes an effective spin of 1/2 representing the
ground state doublet on Tb, and $S^{\mathrm{Pc}_2}$ denotes the real spin of the Pc$_2$ radical. 
The exchange splitting derived from the EPR measurement is $J_\mathrm{eff} = \SI{0.88}{cm^{-1}}$.
Note that the choice of writing the exchange coupling Hamiltonian between two pseudo-spin 1/2 as in
Eq.~(1) implies that \SI{0.88}{cm^{-1}} corresponds to the energy gap between the ground
ferromagnetic exchange Kramers doublet, and the first excited antiferromagnetic exchange Kramers
doublet. It is clear that this small interaction is incompatible with the susceptibility data of
Trojan et al.\cite{Trojan1991, trojan1992strong}

We could find only one other published susceptibility measurement on these systems,
namely for [DyPc$_2]^{0}$.\cite{Branzoli2010} The $\chi T$ data reported by Branzoli et
al.\cite{Branzoli2010} disagree with those of Trojan et al., most significantly in the high
temperature region, where $\chi T$ is substantially higher, reaching a value of 14.6
cm$^3\,\mathrm{K}\,\mathrm{mol}^{-1}$ at 300 K (albeit not fully saturated), compared to 13.0 in
Trojan et al.\ and 14.54 the expected value for the uncoupled system.  

To date, there have been only a few computational studies of the exchange coupling in the
[LnPc$_2]^0$ series. Damjanovi\'c et al., based on a combination of NMR measurements and DFT
calculations, suggested a ferromagnetic interaction between Pc$_2$ radical and Ln(III)
ion.\cite{Damjanovic2015}  However they did not report on the magnitude of the interaction.  DFT
calculations in Ref.\cite{candini2016spin} revealed a correlation between observed magnetic coupling
of [LnPc$_2]^0$ to a Ni surface with computed spin polarization in the Ln 5d orbitals.
Ref.\cite{marocchi2016relay} and the recent work of Pederson et al.\cite{Pederson2019} report
\textit{ab initio} calculations on [TbPc$_2]^0$, similar to those of the present work, but did not
take into account the effect of spin polarization in the ligand $\pi$ system, which we show in the
present work to be important.

This paper presents a theoretical and computational investigation of the
intramolecular exchange coupling mechanisms within [LnPc$_2]^{0}$ (Ln = Tb,
Dy, Ho, and Er) molecules. We find that in the simpler CASSCF calculations
where the active space consists solely of seven Ln$^{3+}$ 4f orbitals and the
singly occupied molecular orbital (SOMO) of the molecular ligands, the coupling
between lanthanides and the radical is constrained to be ferromagnetic by
symmetry, and the exchange strength decreases with increasing atomic number,
i.e., Tb > Dy > Ho > Er. On extension of the active space to RASSCF
calculations, with the previously explored CASSCF active space determining the
RAS2 space, $\pi$ orbitals in RAS1 and $\pi^*$ orbitals in RAS3 space, allowing
for at most double excitations (two holes in the RAS1 space and two particles
in the RAS3 space), a new antiferromagnetic mechanism based on spin
polarization is activated, which reduces the overall exchange coupling constant, 
which remains however ferromagnetic. The computed
exchange splittings are found to be small, of the order of 1--10 \si{cm^{-1}}
for all four ions. Our best value for [TbPc$_2$]$^0$ is $J_\mathrm{eff} = 
\SI{1.92}{cm^{-1}}$, and is consistent in sign and magnitude with the value of 
\SI{0.88}{cm^{-1}} from EPR experiment\cite{komijani2018}, thus supporting 
the interpretation of weak ferromagnetic coupling in [TbPc$_2$]$^0$. These results
are at variance with the strong antiferromagnetic coupling suggested in 
Refs.\ \citenum{Trojan1991, trojan1992strong}. 

\section{Computational Details} 

A series of single point state-averaged CASSCF/RASSCF calculations followed by
state-interaction via spin--orbit coupling (RASSI--SO) was carried out using
the MOLCAS 8.0 code\cite{Aquilante2016}. ANO-RCC-VDZP and ANO-RCC-VDZ basis
sets were used on the lanthanide and the ligand atoms, respectively.  
Crystallographic structures
of [TbPc$_2]^{0}$, [DyPc$_2]^{0}$ and [ErPc$_2]^{0}$ were obtained from the
literature.\cite{katoh2009direct,ostendorp1995phthalocyaninato} The structure
of [HoPc$_2]^{0}$, for which no crystallographic data could be found, was
formed from [TbPc$_2]^{0}$ by replacing Tb with Ho.

CASSCF calculations were performed on [LnPc$_2]^{0}$, employing Cholesky decomposition of the
two-electron integrals (with a threshold of \SI{e-6}{\hartree}).  The CASSCF active space contains
the seven 4f orbitals of the central lanthanide, which transform as $b_2+e_1+e_2+e_3$ in the
approximate D$_\mathrm{4d}$ point group, and the $\pi$-SOMO (having $a_2$
symmetry)\cite{Ishikawa2001} of the Pc$_2$ rings (see Fig.~2, top). RASSCF calculations were
performed on a simplified model structure in order to reduce computational cost. The eight outer
benzene rings of Pc$_2$ were removed and the remaining structure was adapted to perfect
D$_\mathrm{4d}$ symmetry. The resulting structure, referred to as LnPz$_2$ (Pz = porphyrazine), is
shown in Fig.~1. (Cartesian coordinates are given in the Supplementary Information). The same
geometry was used for each of the [LnPz$_2]^0$ considered. All calculations on [LnPz$_2]^0$ were
done without employing Cholesky decomposition of the two-electron integrals. The RAS2 space consists
again of the seven 4f orbitals plus the $\pi$-SOMO. Seven additional occupied $\pi$-MOs are included
in RAS1 (having $a_1+b_1+b_2+e_1+e_3$ symmetries) and four unoccupied $\pi^*$-MOs in RAS3 (having
$e_1+e_3$ symmetries). Up to two holes/particles in RAS1/RAS3 were allowed. This space of 12 active
$\pi$-orbitals was chosen to correspond to the in and out of phase combinations of the 6 frontier
orbitals predicted by a H\"uckel model of the sixteen membered inner ring C$_8$N$_8$ of Pz. These
H\"uckel orbitals have pseudo angular momenta $\lambda=\pm3$, $\pm4$, $\pm5$.

We found that the orbitals of e$_1$ and e$_3$ symmetry had a
tendency to rotate out of the active space. To prevent this from happening,
the 8 orbitals of e$_1$ and e$_3$ symmetry were put into an artificial symmetry
class so as to disable orbital mixing with orbitals outside this class (using
the ``supersymmetry'' keyword of MOLCAS). The validity of this approach relies
on the quality of the starting orbitals. These were obtained from a
state-averaged RASSCF calculation on the twofold degenerate ferromagnetic
($S=7/2$) ground state of [TbPz$_2]^{0}$. This calculation did not experience
the unwanted rotations and provided correct orbitals. 

\begin{figure}
\centering
\includegraphics[scale=.3]{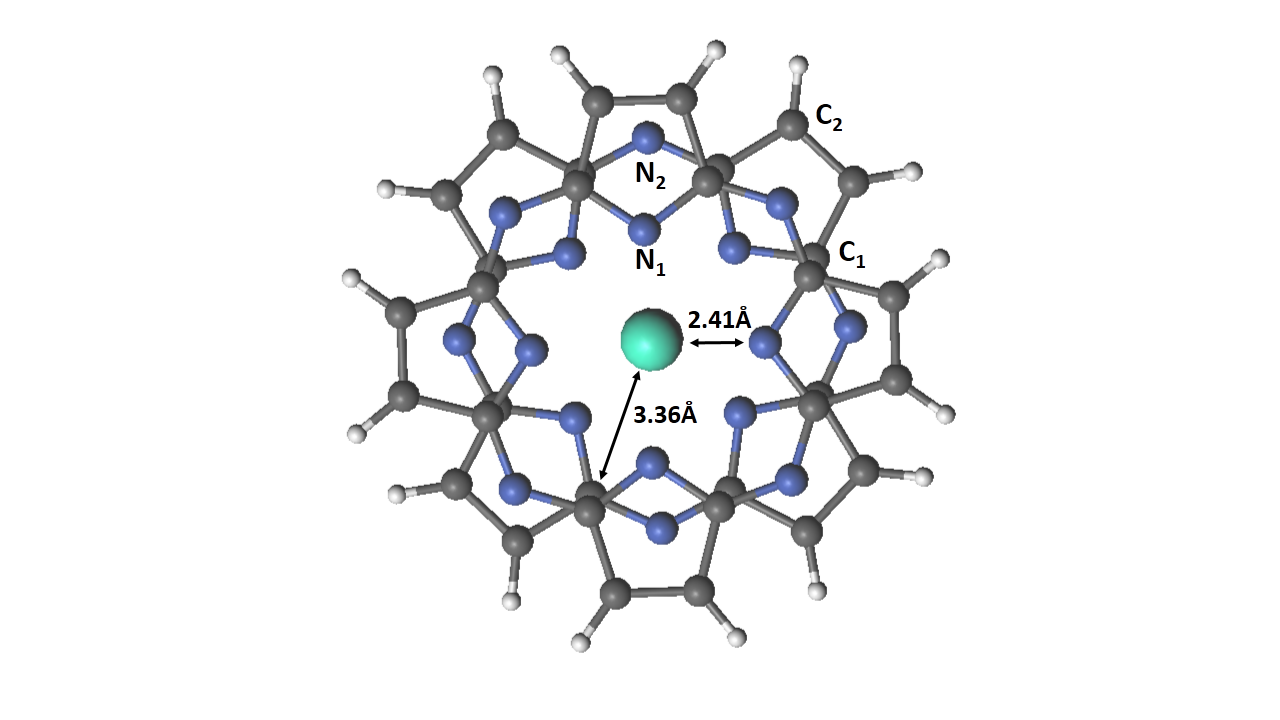}
\caption{Molecular geometry of [LnPz$_2$]$^0$ used in the RASSCF calculations,
where carbon (gray) and nitrogen (blue) atoms are classified into four groups:
C$_1$, C$_2$, N$_1$ and N$_2$. The Ln--N$_1$ and Ln--C$_1$ distances are also
shown.}
\end{figure}

Before spin--orbit coupling (SOC) is considered, the exchange coupling between
the lanthanide and the spin-1/2 radical can be evaluated as the energy
difference between the high-spin and low-spin states that are obtained by
coupling the total spin of the Hund term of the Ln$^{3+}$ ion ($^7$F for
Tb$^{3+}$, $^6$H for Dy$^{3+}$, $^5$I for Ho$^{3+}$, and $^4$I for Er$^{3+}$)
with the spin-1/2 of the radical. In each case, the state-averaging was
performed over all states formally arising from the Hund term.  As an example,
for the [DyPc$_2$]$^{0}$ molecule, we optimize respectively $S=3$ high-spin and
$S=2$ low-spin, with the state average including all 11 spatial components of
the $L=5$ Hund term $^6$H of the Dy$^{3+}$ ion. We then evaluate the exchange
gap  as the difference between the lowest $S=3$
and $S=2$ energies. 

Finally, SOC is introduced by matrix diagonalization in the basis
of all the optimized $S=2$ and $S=3$ CASSCF/RASSCF wavefunctions.

We note that a similar approach was used in a recent computational study of the exchange interaction
in the dimer Ce$_2$(COT)$_3$.\cite{Gendron2019}

\section{Results and discussion}

\begin{figure} [ht!] \centering \includegraphics[width=13.8cm]{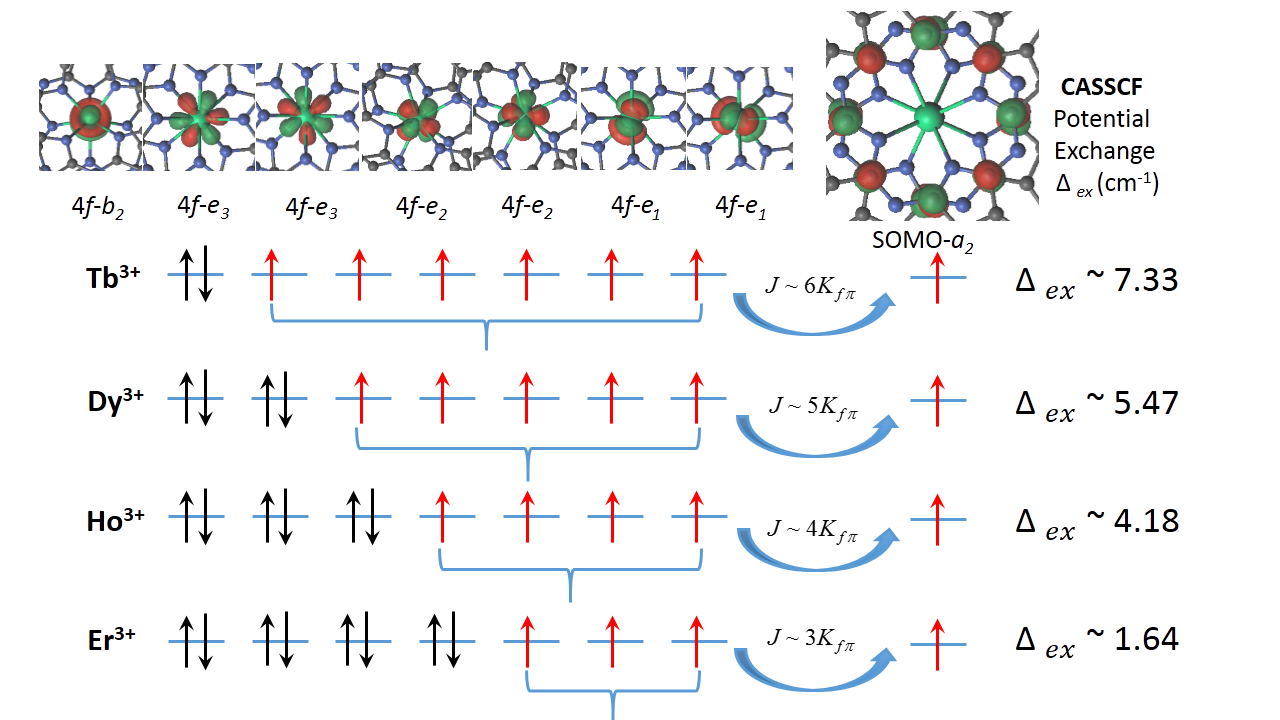} \caption
{Schematic of the ferromagnetic exchange coupling mechanism in the CASSCF
active space of [LnPc$_2]^{0}$. The active space consists of seven Ln
4f orbitals and the radical spin-1/2 orbital as shown on top, with their
symmetry labels in the D$_\mathrm{4d}$ point group.  $K_{f\pi}$ represents a
potential exchange integral between the $\pi$-SOMO and a 4f orbital, and $J$
represents the total exchange strength.} \end{figure} 

The calculated CASSCF active natural orbitals of [LnPc$_2]^{0}$ are shown in
the top of Fig.~2. The Ln 4f orbitals are quasi atomic, while the spin-1/2
radical ($\pi$-SOMO) is mainly localized and evenly distributed on the C$_1$
atoms with nodes on the N atoms.  The exchange gaps obtained from the CASSCF
calculations (without SOC) are listed in Table 1. Our calculations predict
ferromagnetic intramolecular exchange for all four [LnPc$_2]^{0}$ molecules. 

\begin{table}
  \caption{CASSCF exchange gaps (cm$^{-1}$) of [LnPc$_2]^{0}$ without SOC. Positive numbers
  signify ferromagnetic coupling.}
  \label{tbl:Table1}
  \begin{tabular}{cccc}
\hline
 [TbPc$_2]^{0}$ & [DyPc$_2]^{0}$ &[HoPc$_2]^{0}$ &[ErPc$_2]^{0}$\\
\hline
7.33  & 5.47 & 4.18 & 1.64\\
\hline
  \end{tabular}
\end{table}

The occurrence of ferromagnetic exchange interaction in the CASSCF calculations can be
explained on the basis of a symmetry analysis in the approximate
D$_\mathrm{4d}$ point group of the molecule: The SOMO of Pc$_2$ transforms as
$a_2$ (see top right of Fig.~2), while the seven lanthanide 4f orbitals
transform as $b_2+e_1+e_2+e_3$. Thus, the magnetic orbital containing the
Pc$_2$ radical is orthogonal by symmetry to each of the magnetic orbitals of
the lanthanide ion. Kinetic exchange between the magnetic orbitals is
therefore forbidden and only the ferromagnetic potential exchange interaction
is allowed.\cite{anderson1959new,kahn1993molecular} 

Table 1 further shows that the calculated exchange gap decreases from
[TbPc$_2]^{0}$ to [ErPc$_2]^{0}$. This may be understood by considering the
number of unpaired 4f electrons: In the ground Russell--Saunders term of the
Ln$^{3+}$ ions considered here, the number of unpaired 4f electrons decreases
with increasing overall number of 4f electrons (see Fig.~2). If we consider
that each unpaired 4f electron contributes additively to the overall exchange,
the latter is expected to decrease in magnitude in going from [TbPc$_2]^{0}$ to
[ErPc$_2]^{0}$, as observed in Table 1. We note in this respect a recent
experimental work in which the magnetic coupling between TbPc$_2$ and a Ni(111)
surface was also found to decrease along the series Tb--Ni > Dy--Ni >
Er--Ni.\cite{candini2016spin}

\begin{table}[h]
\centering
\caption {CASSCF/RASSI--SO energy levels (cm$^{-1}$) of [LnPc$_2]^{0}$}
\label{tbl:Table2}
\begin{tabular}{ccccc}
\hline 
&[TbPc$_2]^{0}$ & [DyPc$_2]^{0}$  & [HoPc$_2]^{0}$ & [ErPc$_2]^{0}$	\\
\hline 
&0.00   &  0.00  &0.00  & 0.00\\
&0.00   &  0.00  &0.00  & 0.77 \\
&6.18   &  4.01  &3.11  & 0.77\\
&6.18   &  4.01  &3.11  & 1.46 \\
&330.57 & 94.18  &34.26 & 59.62\\
&330.57 & 94.18  &34.26 & 59.96 \\
&335.32 & 98.99  &36.47 & 59.96 \\
&335.32 & 98.99  &36.47 & 60.16\\
&563.61 & 112.11 &52.34 & 155.64\\
&563.61 & 112.11 &52.34 & 155.64\\
&567.34 & 115.87 &55.33 & 155.98 \\
&567.34 & 115.87 &55.33 & 156.56  \\
&...    &    ... &...   & ... \\
\hline 
\multicolumn{5}{c}{g-factors of the two lowest doublets }         \\
\hline
&0.00   & 0.00 & 0.06	& \\
1&0.00   & 0.00 & 0.07	&- \\
&20.00  &19.38 & 20.49	& \\
&0.00   & 0.00 & 0.06	& \\
2&0.00   & 0.00 & 0.07	&- \\
&16.00  &15.37 & 16.44	& \\
\hline  
\end{tabular}
\end{table}

Table 2 lists the lowest lying energy levels of [LnPc$_2]^{0}$ obtained after
diagonalization of the SOC in the CASSCF states. There is a clear separation in
energy between the group of four lowest states and the next group of states. It is
known that the crystal field in these compounds gives rise to a splitting of
the atomic $J$ ground state multiplet of Ln$^{3+}$ into a set of crystal field
levels of which the ground state is a
doublet.\cite{Ishikawa2003determination,Marx2014} Exchange coupling of this
doublet with the spin-1/2 of the radical electron gives rise to the four low-lying
levels seen in Table 2. In the same way, coupling of higher crystal field
levels with the radical results in the groups of higher lying states in Table
2. 

The total exchange splittings in the ground state are seen to be similar in
magnitude to those before SOC, and are again found to decrease, from
\SI{6.18}{cm^{-1}} in [TbPc$_2]^{0}$ to \SI{1.46}{cm^{-1}} in [ErPc$_2]^{0}$.
The effective exchange coupling in the lowest doublet is of Ising type for Tb,
Dy, Ho, but of anisotropic Heisenberg type for Er. We can attribute this
difference in behavior to the different nature of the ground state doublet:
[ErPc$_2]^{0}$ has $M_J=\pm1/2$ as its ground doublet while [TbPc$_2]^{0}$,
[DyPc$_2]^{0}$, [HoPc$_2]^{0}$ all have $|M_J| > 1/2$ ground
states.\cite{Ishikawa2003determination,Marx2014} If we assume that the exchange
between real spins is given by
$-2J\mathbf{S}^\mathrm{Ln}\!\cdot\mathbf{S}^\mathrm{Pc_2}$, projection on
any doublet with $|M_J| > 1/2$ gives an effective Ising coupling:
$-2J_\mathrm{eff}\tilde{S}_z^\mathrm{Ln}S_z^\mathrm{Pc_2}$, while projection on the $M_J=\pm1/2$ doublet
of Er gives an effective anisotropic Heisenberg coupling:
$-2J_\mathrm{eff}(\tilde{S}_z^\mathrm{Ln}S_z^\mathrm{Pc_2}
+8\tilde{S}^\mathrm{Ln}_xS^\mathrm{Pc_2}_x+ 8\tilde{S}_y^\mathrm{Ln}S_y^\mathrm{Pc_2})$, where
$\tilde{S}^\mathrm{Ln}$ is the effective spin of the Ln doublet. Diagonalization of this
anisotropic Heisenberg Hamiltonian gives a spectrum consisting of a
nondegenerate ground state, followed by a doublet at
$8|J_\mathrm{eff}|-J_\mathrm{eff}$, and another nondegenerate state at
$16|J_\mathrm{eff}|$. Referring to Table 2, we observe a qualitative agreement
with the four lowest exchange states of [ErPc$_2]^{0}$.

Establishing the sign of the exchange interaction is not as
straightforward for the calculations with SOC as it is for the calculations
without SOC. For the cases where the exchange is of Ising type (Tb, Dy, Ho) it
can be done by comparing the calculated g-factors of the lowest two doublets
(see Tables 2 and 4). We see that the principal g-factors of the ground doublet
are exactly 4 units higher than those of the next doublet. This corresponds to a
spin flip of the radical electron (whose g-factor is 2), from ferromagnetic
alignment in the ground doublet to antiferromagnetic alignment in the next
doublet. Hence the coupling can be described as ferromagnetic.  This
straightforward interpretation cannot be applied to the case of Er however,
because the exchange is not of Ising type there.
   
\begin{table}
\caption{CASSCF and RASSCF exchange gaps (cm$^{-1}$) of [LnPz$_2]^{0}$ without SOC. Positive numbers
  signify ferromagnetic coupling.}
\begin{tabular}{ccccc}
\hline
& [TbPz$_2]^{0}$ & [DyPz$_2]^{0}$ &[HoPz$_2]^{0}$ &[ErPz$_2]^{0}$\\
\hline
CASSCF	& 7.16  & 5.28 & 4.21 & 2.68\\
RASSCF	& 2.22  & 1.89 & 1.47 & 0.17 \\
\hline
\end{tabular}
\end{table}

We now consider the effect of introducing $\pi$--$\pi^*$ correlation using the
RASSCF method. The calculated values of the exchange gaps before SOC are given
in Table 3. These calculations were done on the smaller model
compounds [LnPz$_2$]$^0$. The absence of the outer benzene rings will affect
the calculated spin density distribution and exchange coupling to some extent,
but we expect that the physics of the exchange mechanisms will be correctly
represented by the LnPz$_2$ models. This is partly confirmed by comparing the CASSCF
values of the exchange gaps in Tables 1 and 3: the sign, order of magnitude and
trend are the same. Further confirmation is provided by the energy levels 
after SOC, which compare well between the LnPz$_2$ (Table S2) and LnPc$_2$ (Table 2) complexes.
Additional computational evidence for the relatively small influence of structural changes on the
low-energy electronic structure of this family of molecules was provided by Pederson et
al.\cite{Pederson2019}

The results in Table 3 show that the RASSCF exchange gaps are still ferromagnetic
but smaller than the corresponding CASSCF gaps. This we interpret as the result of a competition
between a new  \emph{antiferromagnetic} exchange pathway, opened up by activation of $\pi$--$\pi^*$
correlation, and the direct ferromagnetic exchange pathway already present in the CASSCF calculations. 

\begin {table}
\caption {RASSCF/RASSI-SO energy levels (cm$^{-1}$) of [LnPz$_2]^{0}$}
\begin{tabular}{ccccc}
\hline 
&[TbPz$_2]^{0}$ & [DyPz$_2]^{0}$  & [HoPz$_2]^{0}$ & [ErPz$_2]^{0}$	\\
\hline 
&0.00			&0.00  				&0.00 		&0.00\\
&0.00     		&0.00    			&0.00 	    &0.29 \\
&1.92     		&1.84     			&1.22   	&0.29\\
&1.92     		&1.84    			&1.22   	&0.35\\
&325.92   		&84.66    			&26.04		&61.49		 \\
&325.92   		&84.66     			&26.04		&61.49 		\\
&328.14   		&88.70     			&26.47		&61.65 		\\
&328.14   		&88.70     			&26.47		&61.65 		\\
&554.92   		&110.75     		&48.38		&161.46 		\\
&554.92   		&110.75      		&48.38		&161.46		 \\
&556.74   		&112.61    			&49.39		&161.66 		 \\
&556.74   		&112.61    			&49.39		&161.66 		 \\
&...  			&    ...			&	...		&...		 \\
\hline 
\multicolumn{5}{c}{g-factors of the two lowest doublets }         \\
\hline
&0.00   & 0.00 & 0.00	& \\
1&0.00   & 0.00 & 0.00	&- \\
&20.00  &19.37 & 21.97	& \\
&0.00   & 0.00 & 0.00	& \\
2&0.00   & 0.00 & 0.00	&- \\
&16.00  &15.35 & 17.98	& \\
\hline
\end{tabular}
\end{table}

The spin--orbit coupled RASSCF spectrum is given in Table 4. The exchange splittings are smaller
than the corresponding CASSCF values (Table 2 and Table S2) in line with the reduction of the
SOC-free exchange splittings. We note in particular the value for Tb, which decreases from
\SI{6.18}{\cm^{-1}} to {\SI{1.92}{\cm^{-1}}}, closer to the experimental value of
\SI{0.88}{\cm^{-1}}.\cite{komijani2018} 

Note that the recent CASSCF calculations by Pederson et al.\ found $J_\mathrm{eff} =
\SI{8.2}{\cm^{-1}}$ and \SI{6.6}{\cm^{-1}} for two geometries of [TbPc$_2]^0$, which is basically
the same result we obtain with our CASSCF calculation, using an active space where 4f orbitals and
the Pc$_2$ SOMO only are considered. This seems to suggest that the $\pi$--$\pi^*$ excitations
introduce by Pederson et al.\ in their active space were not sufficient to describe the
spin-polarization antiferromagnetic exchange mechanism discovered in this work.

Interestingly, in the Er compound, the relative energies of the four lowest exchange
states cannot be reproduced by a Hamiltonian of the form
$2J_\mathrm{eff}(\tilde{S}_z^\mathrm{Ln}S_z^\mathrm{Pc_2}
+8\tilde{S}^\mathrm{Ln}_xS^\mathrm{Pc_2}_x+ 8\tilde{S}_y^\mathrm{Ln}S_y^\mathrm{Pc_2})$, thus pointing to
a likely breakdown of the 
$2J\mathbf{S}^\mathrm{Ln}\!\cdot\mathbf{S}^\mathrm{Pc_2}$ approximation,
a conclusion also reached in some recent experimental work on Ln--radical exchange
interaction.\cite{Baker2015,Ortu2017}
Recently, Chibotaru, Iwahara, et al.\ have discussed this breakdown on
theoretical grounds using a microscopic model of exchange
interaction.\cite{Chibotaru2015,Iwahara2015,Iwahara2016,Vieru2016} Their model
did not include spin polarization effects on the radical ligand and would thus have
to be extended to be applicable to our case.

It should be noted that an antiferromagnetic exchange coupling pathway, as introduced in the RASSCF
calculations, cannot be explained in terms of interaction between magnetic orbitals on the spin
carriers. We have seen that the SOMO of Pc$_2$/Pz$_2$, belonging to the $a_2$ irrep of
D$_\mathrm{4d}$, is orthogonal by symmetry to the 4f orbitals of Ln$^{3+}$. This absence of orbital
overlap leads to stabilization of the high-spin state, i.e., ferromagnetic
coupling.\cite{kahn1993molecular,Yoshizawa1995}

We attribute this breakdown of the usual model to the effect of spin polarization in the $\pi$
system of the Pc$_2$/Pz$_2$ radical. Spin polarization in radicals of conjugated $\pi$ systems is a
well known effect, and was invoked by McConnell to explain ferromagnetic coupling between stacked
organic radicals (``McConnell's first model'').\cite{mcconnell1963ferromagnetism} Later, Yoshizawa
and Hoffmann argued that these magnetic couplings can be equally well explained on the basis of
interaction between the SOMO's of the organic radicals,\cite{Yoshizawa1995} the condition for
ferromagnetic coupling being again the (near) vanishing of orbital overlap.

Let us now consider the spin density distribution in the Pc$_2$/Pz$_2$ radical.  The SOMO (pictured
in Fig.~2) has amplitudes on the C atoms but nodes on all the N atoms. The spin density, in the
simple molecular orbital picture, is therefore positive on the carbons but zero on the nitrogens.
When we allow for electron correlation in the $\pi$ system (as in our RASSCF calculations), small
but negative spin densities appear on the N atoms. This is illustrated numerically with Mulliken
spin populations in Table 5.

\begin {table} 
\caption{RASSCF Mulliken spin populations $\rho$ on N and C atoms (Fig.1). The
spin-1/2 radical is mainly localized on C$_1$. The small negative spin
populations on the N atoms are due to the spin-polarization effect.}
\begin{tabular}{ccccc }
\hline
Molecules &  $\rho(\mathrm{N}_1)$ &$\rho(\mathrm{N}_2)$ &$\rho(\mathrm{C}_1)$ &
$\rho(\mathrm{C}_2)$ \\
\hline
[TbPz$_2$]$^0$&  -0.0350 & -0.1091 & 0.9468 & 0.2048\\

[DyPz$_2$]$^0$&  -0.0360 & -0.1092 & 0.9466 & 0.2048\\

[HoPz$_2$]$^0$&  -0.0376 & -0.1092 & 0.9462 & 0.2048\\

[ErPz$_2$]$^0$&  -0.0384 & -0.1092 & 0.9460 & 0.2048\\
\hline
\end{tabular}
\end{table}

An elaborate analysis of the interplay between spin polarization and exchange
in [LnPc$_2$]$^0$ will not be attempted here. Instead, a simple argument in the
spirit of McConnell's first model will be given. Let us assume then, that the total
exchange splitting can be estimated as the sum of contributions from each atom
of the ligand, and that only those atoms whose spin populations are non-zero
can contribute. We can also assume that atoms further away from the central
lanthanide ion will have a smaller exchange interaction with it than atoms
closer by. Referring to Fig.~1, the atoms closest to Ln$^{3+}$ are the 8 N$_1$
atoms at 2.41 \AA\ and the 16 C$_1$ atoms at 3.36 \AA. 

In the absence of spin polarization (the CASSCF case) there is only spin
density on C$_1$. Since all C$_1$ atoms are symmetry related, the contribution
from each of them to the exchange interaction must be the same. And since the
overall interaction is ferromagnetic, the contribution from each C$_1$ atom
must be ferromagnetic as well. On the other hand, when spin polarization is
allowed (the RASSCF case), the N$_1$ atoms carry negative spin density, which
will also interact with the lanthanide spin. If we may assume that this
interaction is ferromagnetic, just like that of the C$_1$ atoms, a competition
arises: On the one hand, the majority spin on C$_1$ atoms tries to align itself
parallel to the Ln$^{3+}$ spin, favoring overall ferromagnetic coupling. On the
other hand, the polarized minority spin density on N$_1$ atoms, with an
opposite sign of spin compared with C$_1$ atoms, also tries to be parallel to
the metal spin, favoring overall antiferromagnetic coupling. As a result, the
total exchange interaction is a sum of a positive contribution from C$_1$ and a
negative contribution from N$_1$. Apart from the sign, it is not possible to
determine a priori how large the contribution from N$_1$ is compared to that
from C$_1$. This can be seen from considering the two parameters that will
determine the size of the contribution: the spin density on the atom and the
distance from the atom to the lanthanide ion. The spin density on N$_1$ is
smaller than on C$_1$, but N$_1$ is closer to the lanthanide than C$_1$ (2.41
\AA\ vs.\ 3.36 \AA), so the exchange interaction due to spin density on $N_1$
is stronger than that due to a same amount of spin density on C$_1$. The
resulting contribution from N$_1$ can thus be smaller or larger in absolute
value than the contribution from C$_1$. If it is smaller, the overall exchange
interaction is still ferromagnetic, but weaker than it was before spin
polarization.  On the other hand, if it is larger, the overall exchange
interaction turns from ferromagnetic into antiferromagnetic. In our RASSCF
calculations we observe the first case.

\section{Conclusion}
We have presented results of a computational study of the
intramolecular exchange coupling between Ln$^{3+}$ 4f electrons and
the Pc$_2$ radical in [LnPc$_2]^{0}$ (Ln=Tb, Dy, Ho, and Er) molecules. We
performed a series of state-averaged CASSCF and RASSCF calculations with and
without SOC. When SOC is not considered, CASSCF calculations with minimum
active space show that the coupling between lanthanides and the radical are all
ferromagnetic, and that the magnitude of the exchange gap 
drops as the central metal goes from Tb to Er. On the other hand, inclusion of
additional $\pi$--$\pi^*$ excitations via RASSCF calculations suggests a key
role played by the polarized spin density on the nitrogen atoms, induced by the
spin polarization effect on the Pc$_2$ radical. The negative spin density on
the nitrogen atoms introduces an antiferromagnetic exchange pathway, weakening the 
overall ferromagnetic  coupling strength between lanthanides and the radical.

The small ferromagnetic coupling calculated for [TbPc$_2]^{0}$ agrees with the latest experimental
EPR evidence\cite{komijani2018} but conflicts with the susceptibility measurements of Trojan et
al.\cite{Trojan1991,trojan1992strong} Their data could only be explained by a large
antiferromagnetic coupling, which our calculations do not support.

\begin{acknowledgement}
The authors acknowledge support from the University of Melbourne and the Australian
Research Council (grant ID: DP150103254).  
\end{acknowledgement}

\begin{suppinfo}
Geometry of [LnPz$_2]^0$, CASSCF/RASSI--SO energy levels of [LnPz$_2]^{0}$.
\end{suppinfo}
\bibliography{Article.bib}

\providecommand{\latin}[1]{#1}
\makeatletter
\providecommand{\doi}
  {\begingroup\let\do\@makeother\dospecials
  \catcode`\{=1 \catcode`\}=2 \doi@aux}
\providecommand{\doi@aux}[1]{\endgroup\texttt{#1}}
\makeatother
\providecommand*\mcitethebibliography{\thebibliography}
\csname @ifundefined\endcsname{endmcitethebibliography}
  {\let\endmcitethebibliography\endthebibliography}{}
\begin{mcitethebibliography}{43}
\providecommand*\natexlab[1]{#1}
\providecommand*\mciteSetBstSublistMode[1]{}
\providecommand*\mciteSetBstMaxWidthForm[2]{}
\providecommand*\mciteBstWouldAddEndPuncttrue
  {\def\EndOfBibitem{\unskip.}}
\providecommand*\mciteBstWouldAddEndPunctfalse
  {\let\EndOfBibitem\relax}
\providecommand*\mciteSetBstMidEndSepPunct[3]{}
\providecommand*\mciteSetBstSublistLabelBeginEnd[3]{}
\providecommand*\EndOfBibitem{}
\mciteSetBstSublistMode{f}
\mciteSetBstMaxWidthForm{subitem}{(\alph{mcitesubitemcount})}
\mciteSetBstSublistLabelBeginEnd
  {\mcitemaxwidthsubitemform\space}
  {\relax}
  {\relax}

\bibitem[Woodruff \latin{et~al.}(2013)Woodruff, Winpenny, and
  Layfield]{woodruff2013lanthanide}
Woodruff,~D.~N.; Winpenny,~R.~E.; Layfield,~R.~A. Lanthanide single-molecule
  magnets. \emph{Chemical reviews} \textbf{2013}, \emph{113}, 5110--5148\relax
\mciteBstWouldAddEndPuncttrue
\mciteSetBstMidEndSepPunct{\mcitedefaultmidpunct}
{\mcitedefaultendpunct}{\mcitedefaultseppunct}\relax
\EndOfBibitem
\bibitem[Arom{\'\i} and Brechin(2006)Arom{\'\i}, and
  Brechin]{aromi2006synthesis}
Arom{\'\i},~G.; Brechin,~E.~K. \emph{Single-molecule magnets and related
  phenomena}; Springer, 2006; pp 1--67\relax
\mciteBstWouldAddEndPuncttrue
\mciteSetBstMidEndSepPunct{\mcitedefaultmidpunct}
{\mcitedefaultendpunct}{\mcitedefaultseppunct}\relax
\EndOfBibitem
\bibitem[Wang \latin{et~al.}(2016)Wang, Wang, Bian, Gao, and
  Jiang]{wang2016single}
Wang,~H.; Wang,~B.-W.; Bian,~Y.; Gao,~S.; Jiang,~J. Single-molecule magnetism
  of tetrapyrrole lanthanide compounds with sandwich multiple-decker
  structures. \emph{Coordination Chemistry Reviews} \textbf{2016}, \emph{306},
  195--216\relax
\mciteBstWouldAddEndPuncttrue
\mciteSetBstMidEndSepPunct{\mcitedefaultmidpunct}
{\mcitedefaultendpunct}{\mcitedefaultseppunct}\relax
\EndOfBibitem
\bibitem[Feltham and Brooker(2014)Feltham, and Brooker]{feltham2014review}
Feltham,~H.~L.; Brooker,~S. Review of purely 4f and mixed-metal nd-4f
  single-molecule magnets containing only one lanthanide ion.
  \emph{Coordination Chemistry Reviews} \textbf{2014}, \emph{276}, 1--33\relax
\mciteBstWouldAddEndPuncttrue
\mciteSetBstMidEndSepPunct{\mcitedefaultmidpunct}
{\mcitedefaultendpunct}{\mcitedefaultseppunct}\relax
\EndOfBibitem
\bibitem[Ishikawa \latin{et~al.}(2003)Ishikawa, Sugita, Ishikawa, Koshihara,
  and Kaizu]{ishikawa2003lanthanide}
Ishikawa,~N.; Sugita,~M.; Ishikawa,~T.; Koshihara,~S.-y.; Kaizu,~Y. Lanthanide
  double-decker complexes functioning as magnets at the single-molecular level.
  \emph{Journal of the American Chemical Society} \textbf{2003}, \emph{125},
  8694--8695\relax
\mciteBstWouldAddEndPuncttrue
\mciteSetBstMidEndSepPunct{\mcitedefaultmidpunct}
{\mcitedefaultendpunct}{\mcitedefaultseppunct}\relax
\EndOfBibitem
\bibitem[Ishikawa \latin{et~al.}(2008)Ishikawa, Mizuno, Takamatsu, Ishikawa,
  and Koshihara]{ishikawa2008effects}
Ishikawa,~N.; Mizuno,~Y.; Takamatsu,~S.; Ishikawa,~T.; Koshihara,~S.-y. Effects
  of chemically induced contraction of a coordination polyhedron on the
  dynamical magnetism of bis (phthalocyaninato) disprosium, a single-4f-ionic
  single-molecule magnet with a Kramers ground state. \emph{Inorganic
  chemistry} \textbf{2008}, \emph{47}, 10217--10219\relax
\mciteBstWouldAddEndPuncttrue
\mciteSetBstMidEndSepPunct{\mcitedefaultmidpunct}
{\mcitedefaultendpunct}{\mcitedefaultseppunct}\relax
\EndOfBibitem
\bibitem[Demir \latin{et~al.}(2015)Demir, Jeon, Long, and
  Harris]{demir2015radical}
Demir,~S.; Jeon,~I.-R.; Long,~J.~R.; Harris,~T.~D. Radical ligand-containing
  single-molecule magnets. \emph{Coordination Chemistry Reviews} \textbf{2015},
  \emph{289}, 149--176\relax
\mciteBstWouldAddEndPuncttrue
\mciteSetBstMidEndSepPunct{\mcitedefaultmidpunct}
{\mcitedefaultendpunct}{\mcitedefaultseppunct}\relax
\EndOfBibitem
\bibitem[Vitali \latin{et~al.}(2008)Vitali, Fabris, Conte, Brink, Ruben,
  Baroni, and Kern]{vitali2008electronic}
Vitali,~L.; Fabris,~S.; Conte,~A.~M.; Brink,~S.; Ruben,~M.; Baroni,~S.;
  Kern,~K. Electronic structure of surface-supported bis (phthalocyaninato)
  terbium (III) single molecular magnets. \emph{Nano letters} \textbf{2008},
  \emph{8}, 3364--3368\relax
\mciteBstWouldAddEndPuncttrue
\mciteSetBstMidEndSepPunct{\mcitedefaultmidpunct}
{\mcitedefaultendpunct}{\mcitedefaultseppunct}\relax
\EndOfBibitem
\bibitem[Katoh \latin{et~al.}(2009)Katoh, Yoshida, Yamashita, Miyasaka,
  Breedlove, Kajiwara, Takaishi, Ishikawa, Isshiki, Zhang, Komeda, Yamagishi,
  and Takeya]{katoh2009direct}
Katoh,~K.; Yoshida,~Y.; Yamashita,~M.; Miyasaka,~H.; Breedlove,~B.~K.;
  Kajiwara,~T.; Takaishi,~S.; Ishikawa,~N.; Isshiki,~H.; Zhang,~Y.~F.;
  Komeda,~T.; Yamagishi,~M.; Takeya,~J. Direct observation of lanthanide
  (III)-phthalocyanine molecules on Au (111) by using scanning tunneling
  microscopy and scanning tunneling spectroscopy and thin-film field-effect
  transistor properties of Tb (III)-and Dy (III)-phthalocyanine molecules.
  \emph{Journal of the American Chemical Society} \textbf{2009}, \emph{131},
  9967--9976\relax
\mciteBstWouldAddEndPuncttrue
\mciteSetBstMidEndSepPunct{\mcitedefaultmidpunct}
{\mcitedefaultendpunct}{\mcitedefaultseppunct}\relax
\EndOfBibitem
\bibitem[Stepanow \latin{et~al.}(2010)Stepanow, Honolka, Gambardella, Vitali,
  Abdurakhmanova, Tseng, Rauschenbach, Tait, Sessi, and
  Klyatskaya]{stepanow2010spin}
Stepanow,~S.; Honolka,~J.; Gambardella,~P.; Vitali,~L.; Abdurakhmanova,~N.;
  Tseng,~T.-C.; Rauschenbach,~S.; Tait,~S.~L.; Sessi,~V.; Klyatskaya,~S. Spin
  and orbital magnetic moment anisotropies of monodispersed bis
  (phthalocyaninato) terbium on a copper surface. \emph{Journal of the American
  Chemical Society} \textbf{2010}, \emph{132}, 11900--11901\relax
\mciteBstWouldAddEndPuncttrue
\mciteSetBstMidEndSepPunct{\mcitedefaultmidpunct}
{\mcitedefaultendpunct}{\mcitedefaultseppunct}\relax
\EndOfBibitem
\bibitem[Biagi \latin{et~al.}(2010)Biagi, Fernandez-Rodriguez, Gonidec, Mirone,
  Corradini, Moro, De~Renzi, Del~Pennino, Cezar, and Amabilino]{biagi2010x}
Biagi,~R.; Fernandez-Rodriguez,~J.; Gonidec,~M.; Mirone,~A.; Corradini,~V.;
  Moro,~F.; De~Renzi,~V.; Del~Pennino,~U.; Cezar,~J.; Amabilino,~D. X-ray
  absorption and magnetic circular dichroism investigation of
  bis(phthalocyaninato)terbium single-molecule magnets deposited on graphite.
  \emph{Physical Review B} \textbf{2010}, \emph{82}, 224406\relax
\mciteBstWouldAddEndPuncttrue
\mciteSetBstMidEndSepPunct{\mcitedefaultmidpunct}
{\mcitedefaultendpunct}{\mcitedefaultseppunct}\relax
\EndOfBibitem
\bibitem[Candini \latin{et~al.}(2011)Candini, Klyatskaya, Ruben, Wernsdorfer,
  and Affronte]{candini2011graphene}
Candini,~A.; Klyatskaya,~S.; Ruben,~M.; Wernsdorfer,~W.; Affronte,~M. Graphene
  spintronic devices with molecular nanomagnets. \emph{Nano letters}
  \textbf{2011}, \emph{11}, 2634--2639\relax
\mciteBstWouldAddEndPuncttrue
\mciteSetBstMidEndSepPunct{\mcitedefaultmidpunct}
{\mcitedefaultendpunct}{\mcitedefaultseppunct}\relax
\EndOfBibitem
\bibitem[Rizzini \latin{et~al.}(2011)Rizzini, Krull, Balashov, Kavich, Mugarza,
  Miedema, Thakur, Sessi, Klyatskaya, and Ruben]{rizzini2011coupling}
Rizzini,~A.~L.; Krull,~C.; Balashov,~T.; Kavich,~J.; Mugarza,~A.; Miedema,~P.;
  Thakur,~P.; Sessi,~V.; Klyatskaya,~S.; Ruben,~M. Coupling single molecule
  magnets to ferromagnetic substrates. \emph{Physical review letters}
  \textbf{2011}, \emph{107}, 177205\relax
\mciteBstWouldAddEndPuncttrue
\mciteSetBstMidEndSepPunct{\mcitedefaultmidpunct}
{\mcitedefaultendpunct}{\mcitedefaultseppunct}\relax
\EndOfBibitem
\bibitem[Vincent \latin{et~al.}(2012)Vincent, Klyatskaya, Ruben, Wernsdorfer,
  and Balestro]{vincent2012electronic}
Vincent,~R.; Klyatskaya,~S.; Ruben,~M.; Wernsdorfer,~W.; Balestro,~F.
  Electronic read-out of a single nuclear spin using a molecular spin
  transistor. \emph{Nature} \textbf{2012}, \emph{488}, 357--360\relax
\mciteBstWouldAddEndPuncttrue
\mciteSetBstMidEndSepPunct{\mcitedefaultmidpunct}
{\mcitedefaultendpunct}{\mcitedefaultseppunct}\relax
\EndOfBibitem
\bibitem[Komeda \latin{et~al.}(2014)Komeda, Katoh, and
  Yamashita]{komeda2014double}
Komeda,~T.; Katoh,~K.; Yamashita,~M. Double-decker phthalocyanine complex:
  Scanning tunneling microscopy study of film formation and spin properties.
  \emph{Progress in Surface Science} \textbf{2014}, \emph{89}, 127--160\relax
\mciteBstWouldAddEndPuncttrue
\mciteSetBstMidEndSepPunct{\mcitedefaultmidpunct}
{\mcitedefaultendpunct}{\mcitedefaultseppunct}\relax
\EndOfBibitem
\bibitem[Urdampilleta \latin{et~al.}(2015)Urdampilleta, Klayatskaya, Ruben, and
  Wernsdorfer]{Urdampilleta2015}
Urdampilleta,~M.; Klayatskaya,~S.; Ruben,~M.; Wernsdorfer,~W. Magnetic
  {Interaction} {Between} a {Radical} {Spin} and a {Single}-{Molecule} {Magnet}
  in a {Molecular} {Spin}-{Valve}. \emph{ACS Nano} \textbf{2015}, \emph{9},
  4458--4464\relax
\mciteBstWouldAddEndPuncttrue
\mciteSetBstMidEndSepPunct{\mcitedefaultmidpunct}
{\mcitedefaultendpunct}{\mcitedefaultseppunct}\relax
\EndOfBibitem
\bibitem[Candini \latin{et~al.}(2016)Candini, Klar, Marocchi, Corradini, Biagi,
  De~Renzi, Del~Pennino, Troiani, Bellini, Klyatskaya, Ruben, Kummer, Brookes,
  Huang, Soncini, H, and Affronte]{candini2016spin}
Candini,~A. \latin{et~al.}  Spin-communication channels between Ln (III)
  bis-phthalocyanines molecular nanomagnets and a magnetic substrate.
  \emph{Scientific reports} \textbf{2016}, \emph{6}, 21740\relax
\mciteBstWouldAddEndPuncttrue
\mciteSetBstMidEndSepPunct{\mcitedefaultmidpunct}
{\mcitedefaultendpunct}{\mcitedefaultseppunct}\relax
\EndOfBibitem
\bibitem[Marocchi \latin{et~al.}(2016)Marocchi, Candini, Klar, Van~den Heuvel,
  Huang, Troiani, Corradini, Biagi, De~Renzi, Klyatskaya, Kummer, Brookes,
  Ruben, Wende, del Pennino, Soncini, Affronte, and Bellini]{marocchi2016relay}
Marocchi,~S. \latin{et~al.}  Relay-Like Exchange Mechanism through a Spin
  Radical between TbPc2 Molecules and Graphene/Ni (111) Substrates. \emph{ACS
  nano} \textbf{2016}, \emph{10}, 9353--9360\relax
\mciteBstWouldAddEndPuncttrue
\mciteSetBstMidEndSepPunct{\mcitedefaultmidpunct}
{\mcitedefaultendpunct}{\mcitedefaultseppunct}\relax
\EndOfBibitem
\bibitem[Trojan \latin{et~al.}(1991)Trojan, Hatfield, Kepler, and
  Kirk]{Trojan1991}
Trojan,~K.~L.; Hatfield,~W.~E.; Kepler,~K.~D.; Kirk,~M.~L. Strong exchange
  coupling in lanthanide bis(phthalocyaninato) sandwich compounds.
  \emph{Journal of Applied Physics} \textbf{1991}, \emph{69}, 6007--6009\relax
\mciteBstWouldAddEndPuncttrue
\mciteSetBstMidEndSepPunct{\mcitedefaultmidpunct}
{\mcitedefaultendpunct}{\mcitedefaultseppunct}\relax
\EndOfBibitem
\bibitem[Trojan \latin{et~al.}(1992)Trojan, Kendall, Kepler, and
  Hatfield]{trojan1992strong}
Trojan,~K.~L.; Kendall,~J.~L.; Kepler,~K.~D.; Hatfield,~W.~E. Strong exchange
  coupling between the lanthanide ions and the phthalocyaninato ligand radical
  in bis (phthalocyaninato) lanthanide sandwich compounds. \emph{Inorganica
  chimica acta} \textbf{1992}, \emph{198}, 795--803\relax
\mciteBstWouldAddEndPuncttrue
\mciteSetBstMidEndSepPunct{\mcitedefaultmidpunct}
{\mcitedefaultendpunct}{\mcitedefaultseppunct}\relax
\EndOfBibitem
\bibitem[Kobayashi(2002)]{Kobayashi2002}
Kobayashi,~N. Dimers, trimers and oligomers of phthalocyanines and related
  compounds. \emph{Coordination Chemistry Reviews} \textbf{2002}, \emph{227},
  129--152\relax
\mciteBstWouldAddEndPuncttrue
\mciteSetBstMidEndSepPunct{\mcitedefaultmidpunct}
{\mcitedefaultendpunct}{\mcitedefaultseppunct}\relax
\EndOfBibitem
\bibitem[Dreiser(2015)]{Dreiser2015}
Dreiser,~J. Molecular lanthanide single-ion magnets: from bulk to
  submonolayers. \emph{Journal of Physics: Condensed Matter} \textbf{2015},
  \emph{27}, 183203\relax
\mciteBstWouldAddEndPuncttrue
\mciteSetBstMidEndSepPunct{\mcitedefaultmidpunct}
{\mcitedefaultendpunct}{\mcitedefaultseppunct}\relax
\EndOfBibitem
\bibitem[Komijani \latin{et~al.}(2018)Komijani, Ghirri, Bonizzoni, Klyatskaya,
  Moreno-Pineda, Ruben, Soncini, Affronte, and Hill]{komijani2018}
Komijani,~D.; Ghirri,~A.; Bonizzoni,~C.; Klyatskaya,~S.; Moreno-Pineda,~E.;
  Ruben,~M.; Soncini,~A.; Affronte,~M.; Hill,~S. Radical-lanthanide
  ferromagnetic interaction in a $\mathrm{T}{\mathrm{b}}^{\mathrm{III}}$
  bis-phthalocyaninato complex. \emph{Phys. Rev. Materials} \textbf{2018},
  \emph{2}, 024405\relax
\mciteBstWouldAddEndPuncttrue
\mciteSetBstMidEndSepPunct{\mcitedefaultmidpunct}
{\mcitedefaultendpunct}{\mcitedefaultseppunct}\relax
\EndOfBibitem
\bibitem[Branzoli \latin{et~al.}(2010)Branzoli, Carretta, Filibian, Graf,
  Klyatskaya, Ruben, Coneri, and Dhakal]{Branzoli2010}
Branzoli,~F.; Carretta,~P.; Filibian,~M.; Graf,~M.~J.; Klyatskaya,~S.;
  Ruben,~M.; Coneri,~F.; Dhakal,~P. Spin and charge dynamics in [TbPc$_2$]$^0$
  and [DyPc$_2$]$^0$ single-molecule magnets. \emph{Physical Review B}
  \textbf{2010}, \emph{82}, 134401\relax
\mciteBstWouldAddEndPuncttrue
\mciteSetBstMidEndSepPunct{\mcitedefaultmidpunct}
{\mcitedefaultendpunct}{\mcitedefaultseppunct}\relax
\EndOfBibitem
\bibitem[Damjanovi{\'c} \latin{et~al.}(2015)Damjanovi{\'c}, Morita, Katoh,
  Yamashita, and Enders]{Damjanovic2015}
Damjanovi{\'c},~M.; Morita,~T.; Katoh,~K.; Yamashita,~M.; Enders,~M. Ligand
  $\pi$-Radical Interaction with f-Shell Unpaired Electrons in
  Phthalocyaninato--Lanthanoid Single-Molecule Magnets: A Solution NMR
  Spectroscopic and DFT Study. \emph{Chemistry-A European Journal}
  \textbf{2015}, \emph{21}, 14421--14432\relax
\mciteBstWouldAddEndPuncttrue
\mciteSetBstMidEndSepPunct{\mcitedefaultmidpunct}
{\mcitedefaultendpunct}{\mcitedefaultseppunct}\relax
\EndOfBibitem
\bibitem[Pederson \latin{et~al.}(2019)Pederson, Wysocki, Mayhall, and
  Park]{Pederson2019}
Pederson,~R.; Wysocki,~A.~L.; Mayhall,~N.; Park,~K. Multireference Ab Initio
  Studies of Magnetic Properties of Terbium-Based Single-Molecule Magnets.
  \emph{J. Phys. Chem. A} \textbf{2019}, \emph{123}, 6996--7006, DOI:
  \doi{10.1021/acs.jpca.9b03708}\relax
\mciteBstWouldAddEndPuncttrue
\mciteSetBstMidEndSepPunct{\mcitedefaultmidpunct}
{\mcitedefaultendpunct}{\mcitedefaultseppunct}\relax
\EndOfBibitem
\bibitem[Aquilante \latin{et~al.}(2016)Aquilante, Autschbach, Carlson,
  Chibotaru, Delcey, De~Vico, Fdez.~Galv{\'a}n, Ferr{\'e}, Frutos, Gagliardi,
  Garavelli, Giussani, Hoyer, Li~Manni, Lischka, Ma, Malmqvist, M{\"u}ller,
  Nenov, Olivucci, Pedersen, Peng, Plasser, Pritchard, Reiher, Rivalta,
  Schapiro, Segarra-Mart{\'i}, Stenrup, Truhlar, Ungur, Valentini, Vancoillie,
  Veryazov, Vysotskiy, Weingart, Zapata, and Lindh]{Aquilante2016}
Aquilante,~F. \latin{et~al.}  Molcas 8: {New} capabilities for
  multiconfigurational quantum chemical calculations across the periodic table.
  \emph{Journal of Computational Chemistry} \textbf{2016}, \emph{37},
  506--541\relax
\mciteBstWouldAddEndPuncttrue
\mciteSetBstMidEndSepPunct{\mcitedefaultmidpunct}
{\mcitedefaultendpunct}{\mcitedefaultseppunct}\relax
\EndOfBibitem
\bibitem[Ostendorp \latin{et~al.}(1995)Ostendorp, Werner, and
  Homborg]{ostendorp1995phthalocyaninato}
Ostendorp,~G.; Werner,~J.-P.; Homborg,~H. Bis (phthalocyaninato) erbium
  ($\alpha$1 Phase). \emph{Acta Crystallographica Section C: Crystal Structure
  Communications} \textbf{1995}, \emph{51}, 1125--1128\relax
\mciteBstWouldAddEndPuncttrue
\mciteSetBstMidEndSepPunct{\mcitedefaultmidpunct}
{\mcitedefaultendpunct}{\mcitedefaultseppunct}\relax
\EndOfBibitem
\bibitem[Ishikawa(2001)]{Ishikawa2001}
Ishikawa,~N. Electronic structures and spectral properties of double- and
  triple-decker phthalocyanine complexes in a localized molecular orbital view.
  \emph{Journal of Porphyrins and Phthalocyanines} \textbf{2001}, \emph{05},
  87--101\relax
\mciteBstWouldAddEndPuncttrue
\mciteSetBstMidEndSepPunct{\mcitedefaultmidpunct}
{\mcitedefaultendpunct}{\mcitedefaultseppunct}\relax
\EndOfBibitem
\bibitem[Gendron \latin{et~al.}(2019)Gendron, Autschbach, Malrieu, and
  Bolvin]{Gendron2019}
Gendron,~F.; Autschbach,~J.; Malrieu,~J.-P.; Bolvin,~H. Magnetic Coupling in
  the Ce(III) Dimer Ce$_2$(COT)$_3$. \emph{Inorg. Chem.} \textbf{2019},
  \emph{58}, 581--593, DOI: \doi{10.1021/acs.inorgchem.8b02771}\relax
\mciteBstWouldAddEndPuncttrue
\mciteSetBstMidEndSepPunct{\mcitedefaultmidpunct}
{\mcitedefaultendpunct}{\mcitedefaultseppunct}\relax
\EndOfBibitem
\bibitem[Anderson(1959)]{anderson1959new}
Anderson,~P.~W. New approach to the theory of superexchange interactions.
  \emph{Physical Review} \textbf{1959}, \emph{115}, 2\relax
\mciteBstWouldAddEndPuncttrue
\mciteSetBstMidEndSepPunct{\mcitedefaultmidpunct}
{\mcitedefaultendpunct}{\mcitedefaultseppunct}\relax
\EndOfBibitem
\bibitem[Kahn(1993)]{kahn1993molecular}
Kahn,~O. Molecular magnetism. \emph{VCH Publishers, Inc.(USA), 1993,}
  \textbf{1993}, 393\relax
\mciteBstWouldAddEndPuncttrue
\mciteSetBstMidEndSepPunct{\mcitedefaultmidpunct}
{\mcitedefaultendpunct}{\mcitedefaultseppunct}\relax
\EndOfBibitem
\bibitem[Ishikawa \latin{et~al.}(2003)Ishikawa, Sugita, Okubo, Tanaka, Iino,
  and Kaizu]{Ishikawa2003determination}
Ishikawa,~N.; Sugita,~M.; Okubo,~T.; Tanaka,~N.; Iino,~T.; Kaizu,~Y.
  Determination of ligand-field parameters and f-electronic structures of
  double-decker bis (phthalocyaninato) lanthanide complexes. \emph{Inorganic
  chemistry} \textbf{2003}, \emph{42}, 2440--2446\relax
\mciteBstWouldAddEndPuncttrue
\mciteSetBstMidEndSepPunct{\mcitedefaultmidpunct}
{\mcitedefaultendpunct}{\mcitedefaultseppunct}\relax
\EndOfBibitem
\bibitem[Marx \latin{et~al.}(2014)Marx, Moro, D\"orfel, Ungur, Waters, Jiang,
  Orlita, Taylor, Frey, Chibotaru, and van Slageren]{Marx2014}
Marx,~R.; Moro,~F.; D\"orfel,~M.; Ungur,~L.; Waters,~M.; Jiang,~S.~D.;
  Orlita,~M.; Taylor,~J.; Frey,~W.; Chibotaru,~L.~F.; van Slageren,~J.
  Spectroscopic determination of crystal field splittings in lanthanide double
  deckers. \emph{Chemical Science} \textbf{2014}, \emph{5}, 3287--3293\relax
\mciteBstWouldAddEndPuncttrue
\mciteSetBstMidEndSepPunct{\mcitedefaultmidpunct}
{\mcitedefaultendpunct}{\mcitedefaultseppunct}\relax
\EndOfBibitem
\bibitem[Baker \latin{et~al.}(2015)Baker, Tanaka, Murakami, Ohira-Kawamura,
  Nakajima, Ishida, and Nojiri]{Baker2015}
Baker,~M.~L.; Tanaka,~T.; Murakami,~R.; Ohira-Kawamura,~S.; Nakajima,~K.;
  Ishida,~T.; Nojiri,~H. Relationship between Torsion and Anisotropic Exchange
  Coupling in a Tb$^{\mathrm{III}}$-Radical-Based Single-Molecule Magnet.
  \emph{Inorganic Chemistry} \textbf{2015}, \emph{54}, 5732--5738\relax
\mciteBstWouldAddEndPuncttrue
\mciteSetBstMidEndSepPunct{\mcitedefaultmidpunct}
{\mcitedefaultendpunct}{\mcitedefaultseppunct}\relax
\EndOfBibitem
\bibitem[Ortu \latin{et~al.}(2017)Ortu, Liu, Burton, Fowler, Formanuik, Boulon,
  Chilton, and Mills]{Ortu2017}
Ortu,~F.; Liu,~J.; Burton,~M.; Fowler,~J.~M.; Formanuik,~A.; Boulon,~M.-E.;
  Chilton,~N.~F.; Mills,~D.~P. Analysis of Lanthanide-Radical Magnetic
  Interactions in Ce(III) 2,2'-Bipyridyl Complexes. \emph{Inorganic Chemistry}
  \textbf{2017}, \emph{56}, 2496--2505\relax
\mciteBstWouldAddEndPuncttrue
\mciteSetBstMidEndSepPunct{\mcitedefaultmidpunct}
{\mcitedefaultendpunct}{\mcitedefaultseppunct}\relax
\EndOfBibitem
\bibitem[Chibotaru and Iwahara(2015)Chibotaru, and Iwahara]{Chibotaru2015}
Chibotaru,~L.~F.; Iwahara,~N. Ising exchange interaction in lanthanides and
  actinides. \emph{New Journal of Physics} \textbf{2015}, \emph{17},
  103028\relax
\mciteBstWouldAddEndPuncttrue
\mciteSetBstMidEndSepPunct{\mcitedefaultmidpunct}
{\mcitedefaultendpunct}{\mcitedefaultseppunct}\relax
\EndOfBibitem
\bibitem[Iwahara and Chibotaru(2015)Iwahara, and Chibotaru]{Iwahara2015}
Iwahara,~N.; Chibotaru,~L.~F. Exchange interaction between J multiplets.
  \emph{Phys. Rev. B} \textbf{2015}, \emph{91}, 174438\relax
\mciteBstWouldAddEndPuncttrue
\mciteSetBstMidEndSepPunct{\mcitedefaultmidpunct}
{\mcitedefaultendpunct}{\mcitedefaultseppunct}\relax
\EndOfBibitem
\bibitem[Iwahara and Chibotaru(2016)Iwahara, and Chibotaru]{Iwahara2016}
Iwahara,~N.; Chibotaru,~L.~F. New mechanism of kinetic exchange interaction
  induced by strong magnetic anisotropy. \emph{Scientific Reports}
  \textbf{2016}, \emph{6}, 24743\relax
\mciteBstWouldAddEndPuncttrue
\mciteSetBstMidEndSepPunct{\mcitedefaultmidpunct}
{\mcitedefaultendpunct}{\mcitedefaultseppunct}\relax
\EndOfBibitem
\bibitem[Vieru \latin{et~al.}(2016)Vieru, Iwahara, Ungur, and
  Chibotaru]{Vieru2016}
Vieru,~V.; Iwahara,~N.; Ungur,~L.; Chibotaru,~L.~F. Giant exchange interaction
  in mixed lanthanides. \emph{Scientific Reports} \textbf{2016}, \emph{6},
  24046\relax
\mciteBstWouldAddEndPuncttrue
\mciteSetBstMidEndSepPunct{\mcitedefaultmidpunct}
{\mcitedefaultendpunct}{\mcitedefaultseppunct}\relax
\EndOfBibitem
\bibitem[Yoshizawa and Hoffmann(1995)Yoshizawa, and Hoffmann]{Yoshizawa1995}
Yoshizawa,~K.; Hoffmann,~R. The role of orbital interactions in determining
  ferromagnetic coupling in organic molecular assemblies. \emph{Journal of the
  American Chemical Society} \textbf{1995}, \emph{117}, 6921--6926\relax
\mciteBstWouldAddEndPuncttrue
\mciteSetBstMidEndSepPunct{\mcitedefaultmidpunct}
{\mcitedefaultendpunct}{\mcitedefaultseppunct}\relax
\EndOfBibitem
\bibitem[McConnell(1963)]{mcconnell1963ferromagnetism}
McConnell,~H.~M. Ferromagnetism in solid free radicals. \emph{The Journal of
  Chemical Physics} \textbf{1963}, \emph{39}, 1910\relax
\mciteBstWouldAddEndPuncttrue
\mciteSetBstMidEndSepPunct{\mcitedefaultmidpunct}
{\mcitedefaultendpunct}{\mcitedefaultseppunct}\relax
\EndOfBibitem
\end{mcitethebibliography}

\end{document}


\section{Geometry of [LnPz$_2]^{0}$ molecules}

\begin {table} 
\begin{tabular}{ c c c c}
\hline 
D$_\mathrm{4d}$-[LnPz$_2]^{0}$   & X & Y & Z\\
\hline 
Ln   &  0.00000  &	0.00000&0.00000\\
N   & 1.96363   &0.00000	&1.39550\\
N   &  2.39530  &2.39530	&1.56000\\
C   & 2.77855   &1.10129	&1.53290\\
C   & 4.17999   &0.70300	&1.66060\\
H   &  4.99972  &1.32549	&1.73529\\   
\hline    
\end{tabular}
\caption {Cartesian coordinates (\AA) of the D$_\mathrm{4d}$ symmetry unique atoms in 
[LnPz$_2]^{0}$.}
\end {table}

\section{CASSCF/RASSI--SO energies of [LnPz$_2]^{0}$}

\begin {table}
\centering
\begin{tabular}{ccccc}
\hline 
&[TbPz$_2]^{0}$ & [DyPz$_2]^{0}$  & [HoPz$_2]^{0}$ & [ErPz$_2]^{0}$	\\
\hline 
&0.00   &  0.00	& 	0.00		&0.00\\
&0.00   &  0.00	&	0.00		&0.98 \\
&6.09   		&  3.88    		&	3.33		&0.98		\\
&6.09   		&  3.88    		&	3.33		&2.20		\\
&325.82   		&  86.40  		&	24.99		&62.09		 \\
&325.82   		&  86.40   		&	24.99		&62.09 		\\
&330.53   		&  91.27   		&	26.95		&62.48 		\\
&330.53   		&  91.27   		&	26.95		&62.48 		\\
&554.47   		&  110.01   	&	48.31		&161.59 		\\
&554.47   		&  110.01    	&	48.31		&161.59		 \\
&558.03   		&  113.27  		&	50.94		&162.38 		 \\
&558.03   		&  113.27  		&	50.94		&162.38 		 \\
&    ...		&	...			&...			&...		 \\
\hline 
\multicolumn{5}{c}{g-factors of the two lowest doublets }         \\
\hline
&0.00   & 0.00 & 0.00	& \\
1&0.00   & 0.00 & 0.00	&- \\
&20.00  &19.36 & 21.97	& \\
&0.00   & 0.00 & 0.00	& \\
2&0.00   & 0.00 & 0.00	&- \\
&16.00  &15.35 & 17.97	& \\
\hline
\end{tabular}
\caption{CASSCF/RASSI--SO energy levels (cm$^{-1}$) of [LnPz$_2]^{0}$}
\end {table}


\clearpage
